\documentclass[manuscript]{acmart}
\usepackage{stfloats}

\AtBeginDocument{%
  \providecommand\BibTeX{{%
    \normalfont B\kern-0.5em{\scshape i\kern-0.25em b}\kern-0.8em\TeX}}}

\copyrightyear{2023}
\acmYear{2023}
\setcopyright{rightsretained}
\acmConference[RecSys '23]{Seventeenth ACM Conference on Recommender Systems}{September 18--22, 2023}{Singapore, Singapore}
\acmBooktitle{Seventeenth ACM Conference on Recommender Systems (RecSys '23), September 18--22, 2023, Singapore, Singapore}\acmDOI{10.1145/3604915.3608863}
\acmISBN{979-8-4007-0241-9/23/09}

\begin{document}

\title[LightSAGE: GNN Item Retrieval at Shopee Ads Recommendation]{LightSAGE: Graph Neural Networks for Large Scale Item Retrieval in Shopee's Advertisement Recommendation}

\author{Dang Minh Nguyen}
\authornote{Both authors contributed equally to this research.}
\email{dangminh.nguyen@shopee.com}
\orcid{0000-0002-5951-7229}
\affiliation{%
  \institution{Shopee, SEA Group}
  \streetaddress{5 Science Park Dr}
  \country{Singapore}
  \postcode{118265}
}

\author{Chenfei Wang}
\authornotemark[1]
\email{chenfei.wang@shopee.com}
\orcid{0009-0004-8423-1136}
\affiliation{%
 \institution{Shopee, SEA Group}
  \streetaddress{5 Science Park Dr}
  \country{Singapore}
  \postcode{118265}
  }

\author{Yan Shen}
\email{sheny@sea.com}
\orcid{0009-0000-5247-4150}
\affiliation{%
 \institution{Shopee, SEA Group}
  \streetaddress{5 Science Park Dr}
  \country{Singapore}
  \postcode{118265}
  }

\author{Yifan Zeng}
\email{alan.zeng@shopee.com}
\orcid{0009-0000-6143-3043}
\affiliation{%
 \institution{Shopee, SEA Group}
  \city{Beijing}
  \country{China}
  }

\renewcommand{\shortauthors}{DM Nguyen, C Wang, Y Shen et.al.}

\begin{abstract}
  Graph Neural Network (GNN) is the trending solution for item retrieval in recommendation problems. Most recent reports, however, focus heavily on new model architectures. This may bring some gaps when applying GNN in the industrial setup, where, besides the model, constructing the graph and handling data sparsity also play critical roles in the overall success of the project. In this work, we report how GNN is applied for large-scale e-commerce item retrieval at Shopee. We introduce our simple yet novel and impactful techniques in graph construction, modeling, and handling data skewness. Specifically, we construct high-quality item graphs by combining strong-signal user behaviors with high-precision collaborative filtering (CF) algorithm. We then develop a new GNN architecture named LightSAGE to produce high-quality items' embeddings for vector search. Finally, we design multiple strategies to handle cold-start and long-tail items, which are critical in an advertisement (ads) system. Our models bring improvement in offline evaluations, online A/B tests, and are deployed to the main traffic of Shopee's Recommendation Advertisement system. 
\end{abstract}




\maketitle

\section{Introduction}

In e-commerce platforms like Shopee, a classic Recommendation System (RS) problem is: from a pool of billions of candidate items, retrieve those that are similar to users' currently viewing items. For such an item-to-item problem, the trending solution is to use GNNs on an item graph to obtain items' embeddings, then perform retrieval using approximate nearest neighbor search (ANNs). Prior arts \cite{wang2019ngcf,hamilton2017graphsage,he2020lightgcn,ying2018pinsage,wu2021self,wang2019hgat,wang2019kgat,wu2022graph,gao2023survey} have reported on the potential of GNN in RS. However, most of them focus heavily on the model architecture. Applying GNN in a large-scale e-commerce system faces three critical challenges:
\begin{itemize}
    \item \textbf{Graph construction}: Most existing works construct the item graphs by connecting items consecutively clicked by the same users within a short time window \cite{wang2018eges, vasile2016meta, barkan2016item2vec}. However, since e-commerce landing pages are usually filled with a wide range of products to meet users' diverse needs, user behaviors contain lots of noise and even successive clicks can come from unrelated items.
    \item \textbf{Model architecture}: Most popular GNN architectures cannot be used off-the-shelf for Shopee. PinSAGE \cite{ying2018pinsage} contains burdensome components, namely feature transformation and linear activation, which may deteriorate the model's performance. Meanwhile, LightGCN \cite{he2020lightgcn} and GAT \cite{velivckovic2017gat} are too memory and computationally expensive.
    \item \textbf{Sparsity and long-tail data}: The users-items interactions are highly skewed. A small set of popular items, referred to as "seed items", gets most of the interactions, from which GNN can learn their embedding well. The majority, however, are either new or rarely clicked and are referred to as "long-tail items". They do not show up in the training graph and GNN model cannot learn them. An ads system cannot ignore these items as it will hurt the seller's ads engagement.
    
\end{itemize}
In this work, we present how we overcome these challenges to implement GNN-based item-to-item retrieval at Shopee's Advertisement Recommendation system. First, we prioritize precision during graph construction by considering only strong-signal user-item interactions together with high-precision CF algorithms. Second, we propose our own GNN architecture named LightSAGE, which is inspired by PinSAGE's neighborhood sampling \cite{ying2018pinsage} and LightGCN's embedding propagation and aggregation \cite{he2020lightgcn}. Third, from the GNN-trained seed embeddings, we develop multiple strategies leveraging content features and the latest user behaviors to populate embeddings for all items. The impacts of our techniques are tested with both offline evaluation and online A/B tests.

\section{Methodology}

\subsection{Graph Construction}
Our focus is on the similarities among items. Therefore, we design our graph to be a homogeneous directed graph, with items as the only node type. The node's features are the item's contents features, such as id, product category, price, text embedding, etc. There are no user nodes, instead, the user's behaviors are encoded in the graph edges.

\begin{figure*}[htbp]
    \centering
    \includegraphics[scale=0.425]{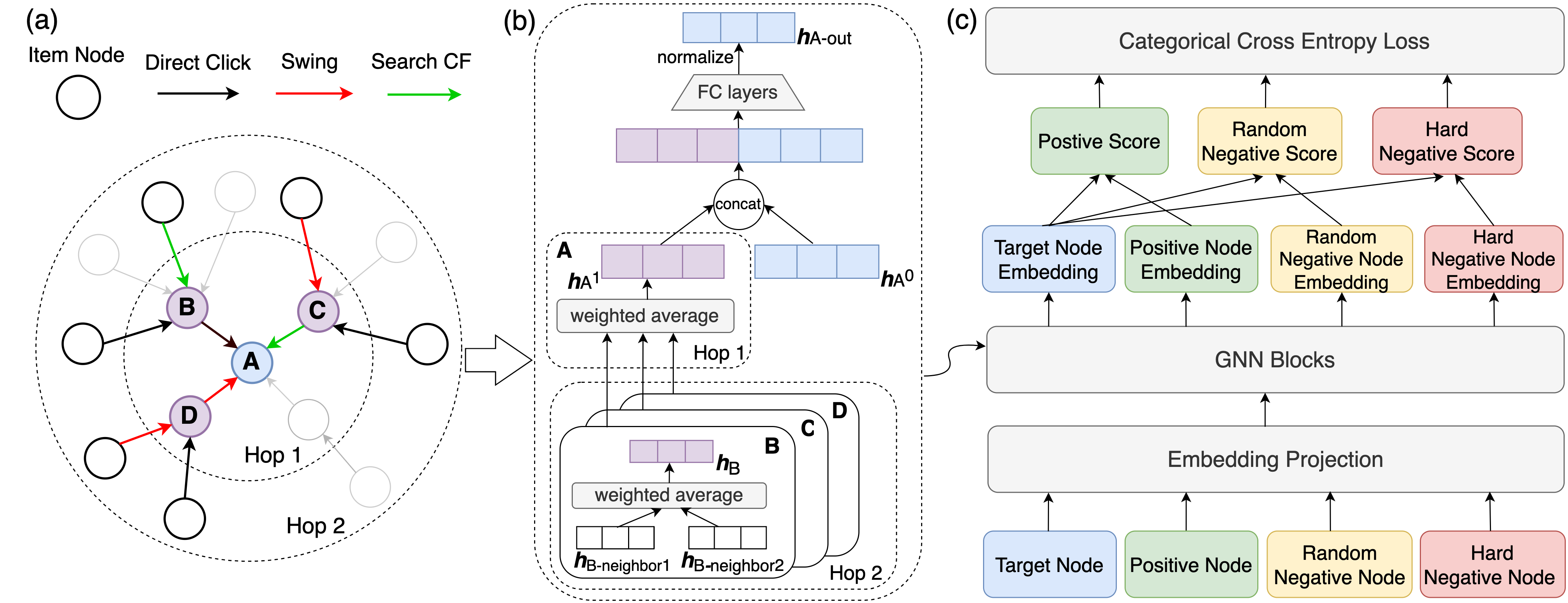}
    \caption{Overview design of our GNN: (a) Logic of edge formation and neighbor sampling (b) Aggregation logic of the GNN block model (c) The overview architecture and training flow of LightSAGE}
    \label{fig:model_architect}
    \Description[(a) Graph edges from user's direct click, swing and search cf algorithm (b) The neighbors' embeddings are aggregated, then combined with central embeddings to form next layer embeddings (c) Model components, which have projection layer, GNN blocks, and cross-entropy loss on positive/negative embedding]{(a) Graph edges are formed from the user's direct click, swing and search cf algorithm (b) GNN block, in which the neighbors' embeddings are aggregated, then combined with central embeddings and pass through the FC-layer to form next layer embeddings (c) Model components, each positive/negative node passes through embedding projection layer, then through GNN blocks to produce embeddings, which are then used to calculate cosine-similarity score for cross-entropy loss}
\end{figure*}

Figure \ref{fig:model_architect}a depicts our logic for constructing the edges. To achieve the best quality for the seed embeddings, we prioritize precision and only consider the user's behaviors in a specific scenario. When a user is at the Product Detail Page of item B, he will be presented with other items that may be of his interest. If the user clicks on item A, it reflects his perspective that these items are highly related, and we form an edge from B to A. The weight of this edge represents the number of users sharing such behaviors. We filter out behaviors from spam users and weak edges with very low click counts to eliminate noises.

Limiting ourselves to a specific scenario makes our graph sparse. To increase the graph density, we add more edges based on high-precision item-to-item algorithms: if given item C, the algorithms recommend item A, then an edge is formed from C to A. The weight of this edge is calculated by converting the algorithm confidence scores into users' click counts with some penalties to reflect its lower priority as compared to actual direct clicks. Any high-precision item-to-item algorithm can be used as supplementary links. In this work, we use two CF algorithms that have proven their success in item retrieval at Shopee:
\begin{itemize}
    \item \textbf{Swing}: an implementation of Swing algorithm \cite{yang2020swing}, which finds substitutable items based on the substructures of user-item click bi-partitive graph.
    \item \textbf{Search CF}: an algorithm based on users' search - another strong-signal behavior: if the same user searches the same keywords, then clicks on both item A and B, then A can recommend B and vice-versa.
\end{itemize}

\subsection{Model Training}
The overall model architecture is illustrated in Fig. \ref{fig:model_architect}b and \ref{fig:model_architect}c. During training, the model inputs are dynamically sampled from the graph. Each training sample contains one target node, one positive node and multiple negative nodes. For each item in the graph, i.e. target node, we randomly pick a positive node from its direct neighbors. There are two types of negative nodes: random negative nodes, generated by degree-based negative sampling \cite{mikolov2013distributed}, and hard negative nodes, generated dynamically using in-batch hard negative sampling \cite{huang2020embedding}. We sample $k$ layers of neighbors for each node in the training sample using random walk, similar to PinSAGE \cite{ying2018pinsage}. The nodes and their sub-graph neighbors are then fed into the model as one batch.

A node has both sparse, dense, and pre-trained features, which are converted into dense vectors in the Projection layer and combined to form layer 0 embeddings. The GNN Blocks propagate and aggregate the current layer's embedding of the neighboring nodes to create the next layer embeddings of the center nodes. As shown in Fig. \ref{fig:model_architect}b, the embedding $h_{A-out}$ of the center node A is produced from its current embedding $h_A^0$ and the aggregated embedding $h_A^1$ obtained by a 2-hop propagation of its neighbors B, C and D. The aggregation logic is similar to LightGCN \cite{he2020lightgcn}, with feature transformation and nonlinear activation removed. The final embeddings are then used to calculate the cosine-similarity scores between the target node and the positive/negative nodes. The model is trained with classification tasks using cross-entropy loss on these scores.

\subsection{Handling Long-tail Items}
The final output of the GNN model is the embeddings of the seed items. We rely on these seed values to populate embeddings for all other long-tail items using two main logic. The first logic makes use of the items' contents. For each long-tail item, we find seed items having similar content features (category, price, brand, etc.) and average their embeddings to represent that long-tail item. This logic alone can theoretically populate for $100\%$ of the item pool. However, it cannot capture users' behaviors that take place after model training, thus losing valuable information and being vulnerable to data drift. 

The second logic addresses the pitfalls of the first logic by introducing an inference graph. The inference graph is built using the same method as the training graph but with more relaxed criteria as our focus shifts from precision to recall. The graph is rebuilt daily to capture the latest items pool and user behaviors. If a long-tail item has seed items as its neighbors, its embedding is derived by weighted averaging embeddings of the neighboring seeds. This logic is equivalent to the aggregation step of LightGCN \cite{he2020lightgcn}. Embeddings produced this way carry information on the latest user behaviors, and have better quality as compared to the first logic. We use both logic in conjunction. A seed item retains its embedding from the GNN model. A long-tail item presented in the inference graph will follow logic two, otherwise, it will follow logic one.

\section{Experiments}

\subsection{Offline Evaluations}

Our offline experiments are conducted using real user behavior data collected from the Shopee app. Training data used to construct the graph is sampled from users' clicks in one of Shopee's markets in 30 days between 2023 Feb and 2023 Mar. It involves 30 million unique users, 30 million unique products, and 3 billion clicks. The long-tail items are those whose clicks belong to the bottom 90 percent of the item pools.

\begin{table*} [htbp]
  \caption{Offline performance of LightSAGE as compared to other architectures.}
  \label{tab:model_improvement}
  \begin{tabular}{cccc}
Model           & AUC               & Unique recall     & Tail unique recall \\ 
\midrule
Node2vec        & 0.872               & +0.00\%           & +0.00\%                  \\
PinSAGE       & 0.882 (+1.02\%)             & +0.81\%           & +0.29\%                   \\
GAT             & 0.878 (+0.62\%)            & +0.79\%           & -1.44\%               \\
LightSAGE       & \textbf{0.890 (+1.92\%)}   & \textbf{+1.42\%}  & \textbf{+1.29\%}              \\          
\end{tabular}
\end{table*}

\begin{table*}[htbp]
  \caption{Impact of improvement made in terms of model, graph construction, and long-tail handling.}
  \label{tab:ablation_study}
  \begin{tabular}{ccccc}
Model       & Graph                 & Tail handling     & Unique recall         & Tail unique recall \\
\midrule
Node2vec    & Click sequences       & No                & +0.00\%               & +0.00\%           \\
LightSAGE   & Click sequences       & No                & + 0.99\%              & +0.20\%              \\
LightSAGE   & Direct clicks         & No                & +70.20\%              & +13.45\%               \\
LightSAGE   & Direct clicks + CF    & No                & +81.90\%              & +112.37\%              \\
LightSAGE   & Direct clicks + CF    & Content           & +74.39\%              & +235.93\%              \\
LightSAGE   & Direct clicks + CF    & Content + graph   & \textbf{+88.40\%}     & \textbf{+303.29\%}              \\
\end{tabular}
\end{table*}

The experiments are evaluated using two metrics. The first metric is \textbf{AUC} on the link prediction task. The model is trained on 95\% of the graph links, then is tasked with predicting if another node is its neighbor. Here the positive samples are the left-out 5\% of the true links and negative samples are in-batch hard negatives.

The second metric is \textbf{unique recall rate} on future users' clicks. Using the item's embeddings, we perform ANNs to simulate retrieval on the same market and compare the results against users' clicks recorded in the following days. The recall metric is similar to Ref \cite{yang2020swing}, which measures the overlapping between the retrieved items and the user's future clicks. Here, we make 2 modifications. First, since in production, LightSAGE is used in conjunction with other retrieval algorithms like Swing \cite{yang2020swing}, we focus on the unique recall rate, which considers items retrieved by only LightSAGE and not others. Second, since we are interested in long-tail items, we calculate another metric named \textbf{tail unique recall}, using only data involving long-tail items.

Table \ref{tab:model_improvement} shows offline metrics of different model architectures when trained on our best graphs without long-tail handling logic. The recall metrics are normalized by taking relative values as compared to the Node2vec model. It can be seen that LightSAGE has the best performance in both AUC, unique recall and tail unique recall among our chosen baselines.

Table \ref{tab:ablation_study} shows the results of the ablation study, where each improvement is incrementally added. With the click-sequence graph, LightSAGE performs slightly better than Node2vec. When the graph is switched to direct clicks, both the general and tail recall rate see enormous improvement. The addition of CF links and tail-handling techniques further boosts the metrics, especially for long-tail items. Graph-wise, CF links help increase the node counts by 66\% and edge counts by 177\%. The best performance is achieved when all the logic improvements are present, i.e. LightSAGE model trained using graphs from direct clicks and CF algorithms, then assisted with long-tail handling techniques during inference.

\subsection{Online A/B Test}
We conduct online A/B experiments by deploying LightSAGE and related logic to the Product Detail Page (PDP) section in Shopee's app. The control group has all existing retrieval strategies in our production system. The treatment group incorporates LightSAGE in addition to those same strategies. Both groups have the same ranking and business logic for a fair comparison.

Overall, LightSAGE increases the ads orders by 1\% and revenue by 5\%. The revenue boost is due to the increase in the number of retrieved relevant items, which intensifies the ads bidding auction. LightSAGE also promotes the healthiness of the platform, as long-tail ads have more chances to reach users. The number of unique ads that have users' impressions and clicks increase by 10\% and 3\%, respectively. As a result of this significant improvement, LightSAGE is deployed to the main traffic of Shopee's PDP in all markets.

\section{Conclusions}
In this work, we present our methods to apply GNN for large-scale e-commerce items retrieval at Shopee's ads RS. We detail novel techniques in graph construction, model training, and long-tail item handling. The offline evaluation shows the positive impacts of our improvement on AUC, unique recall rate, and tail recall rate. Online A/B test confirmed the positive impacts on business metrics and the healthiness of the platform.

\begin{acks}
We thank Marcus Tan Yi Xiang for his contribution to the code development, and Weijie Bian for his help in reviewing the manuscript.
\end{acks}

\bibliographystyle{ACM-Reference-Format}
\bibliography{bib}

\appendix

\end{document}